\documentclass[prl,a4paper,superscriptaddress,10pt,twocolumn,showpacs,showkeys,titlepage]{revtex4}

\usepackage{times}
\usepackage{amsmath}
\usepackage{amssymb}
\usepackage[dvipdf]{graphicx}
\usepackage{hyperref}

\begin{document}

\title[Partial Teleportation of entangled atomic states]{Partial
teleportation of entangled atomic states}
\author{W. B. Cardoso}
\email{wesleybcardoso@gmail.com}
\affiliation{Instituto de F\'{\i}sica, Universidade Federal de Goi\'{a}s, 74.001-970, Goi%
\^{a}nia (GO), Brazil.}
\author{N. G. de Almeida}
\affiliation{N\'{u}cleo de Pesquisas em F\'{\i}sica, Universidade Cat\'{o}lica de Goi\'{a}%
s, 74.605-220, Goi\^{a}nia (GO), Brazil.}

\begin{abstract}
In this paper we propose a scheme for partially teleporting entangled atomic
states. Our scheme can be implemented using only four two-level atoms
interacting either resonantly or off-resonantly with a single cavity-QED.
The estimative of losses occurring during this partial teleportation process
is accomplished through the phenomenological operator approach technique.
\end{abstract}

\maketitle

\section{Introduction}

Quantum teleportation, first suggested by Bennett \textit{et al.} \cite%
{Bennett93}, is one of the cornerstones of quantum information and
computation \cite{Brassard98,Gottesman99,Nielsen00}. The crucial ingredient
characterizing this phenomenon is the transfer of information between
noninteracting systems at the expense of a quantum channel. This issue has
received great attention since its pioneer proposal, mainly after its
experimental realizations from 1997 onwards \cite%
{Bouwmeester97,Boschi98,experiments,experiments2,experiments3}. In the
meantime, various proposals have been suggested for implementing
teleportation, for instance, teleportation of trapped wave fields inside
high-Q microwave cavities \cite%
{Davidovich94,Cirac94,Almeida00,Pires04,Yang06,Cardoso08}, teleportation of
running wave fields \cite{BraunsteinPRL98,Villas-BoasPRA99,LeePRA00},
teleportation of trapped field states inside a single bimodal cavity \cite%
{Iara07}, nonprobabilistic teleportation of a field state via cavity QED
\cite{Garreau07}, and teleportation of the angular spectrum of a
single-photon field \cite{Matos07}, among others.

Since the pioneering work by Bennet \textit{et al.} \cite{Bennett93},
several schemes for teleportation differing from this original protocol (OP)
have appeared in the literature. For example, in Ref. \cite{LeePRA00} the
authors show how to partially teleport an entanglement of zero and one
photon state in the running wave domain. By partial teleportation (PT) it is
mean that teleportation will occur by changing one of the partners of the
entangled state to be teleported. PT can be detailed, step by step, in the
following sequence: i) particles $A$ and $B$ are previously prepared in the
state that we want to teleport; ii) an entangled state of particles $C$ and $%
D$ is created; iii) a joint measurement on particles $B$ and $C$ is
accomplished such that particles $A$ and $D$ assumes the previous entangled
state from particles $A$ and $B$. Note that the term PT is also used in
literature to deal with teleportation of an unknown state with the
generation of its clone \cite{FilipPRA04,ZhaoPRL05}. Other interesting
protocol is the so called entanglement swapping \cite%
{ZukowskiPRL93,BosePRA98,LuPLA00}. In a standard entanglement swapping,
there are usually three spatially separate users: Alice, Bob, and Charlie.
Alice and Bob share pairs of entangled particles with Charlie. Initially,
the particles with Alice and Bob are not entangled. Then, Charlie makes a
Bell-state measurement on his two particles, leading to the entanglement of
the two particles with Alice and Bob.

In this paper we present a scheme for partial teleportation in the cavity
QED domain using entanglement swapping in only one particle, as is done in
Ref. \cite{LeePRA00}. Our scheme uses four two-level atoms interacting
either on or off resonantly with a single mode of a high-Q cavity, Ramsey
zones, and selective atomic state detectors. To estimate losses occurring
during the partial teleportation process, we used the phenomenological
operator approach technique (POA) \cite{poa}.

\section{Ideal Teleportation Process}

We assume atom $1$ previously entangled with atom $2$ in the following
state, which is the state we want to teleport
\begin{equation}
\left\vert \phi \right\rangle _{12}=C_{0}\left\vert g\right\rangle
_{1}\left\vert e\right\rangle _{2}+C_{1}\left\vert e\right\rangle
_{1}\left\vert g\right\rangle _{2}\text{,}  \label{in}
\end{equation}%
where $C_{0}$ and $C_{1}$ are unknown coefficients obeying $\left\vert
C_{0}\right\vert ^{2}+\left\vert C_{1}\right\vert ^{2}=1$, and $\left\vert
g\right\rangle $ ($\left\vert e\right\rangle $) is the atomic ground
(excited) state. The state (\ref{in}) can be prepared, for instance, by the
method presented in Ref. \cite{zhengPRL00}, where two two-level atoms
interact simultaneously with a single mode of a cavity-field.

The Hamiltonian describing the atom-field interaction, in the interaction
picture, is
\begin{equation}
H_{I}=\hbar \lambda \left( a^{\dagger }\sigma ^{-}+a\sigma ^{+}\right),
\label{hon}
\end{equation}
when the atom-field interaction is resonant, and
\begin{equation}
H_{I}=\frac{\hbar\lambda^{2}}{\delta}a^{\dagger}a \sigma^{+}\sigma^{-},
\label{hid}
\end{equation}
when the atom-field interaction is off-resonant. This condition is valid
provided that $\overline{n}\lambda ^{2}\ll \delta ^{2}$ $+\gamma ^{2}$,
where $\overline{n}$ is the mean photon number and $\gamma $ is the damping
rate for the cavity-field. Here $a^{\dagger }$ and $a$ are the creation and
annihilation operators for the cavity field mode, $\sigma ^{+}$ and $\sigma
^{-}$ are the raising and lowering operators for the atom, $\lambda $ is the
atom-field coupling constant, and $\delta =\omega -\omega _{0}$ is the
detuning between the cavity field frequency $\omega $ and the atomic
frequency $\omega _{0}$.

To compose the nonlocal channel, a third atom (initially prepared in the
excited state $|e\rangle _{3}$) interacts resonantly with the cavity field
mode $A$ (in vacuum state $|0\rangle _{A}$), according to Eq. (\ref{hon})
(see Fig. 1a). Adjusting the atom-field interaction time to $t=\pi /4\lambda
$, the nonlocal channel will be given by%
\begin{equation}
\left\vert \psi \right\rangle =\frac{1}{\sqrt{2}}\left( \left\vert
e\right\rangle _{3}\left\vert 0\right\rangle _{A}-i\left\vert g\right\rangle
_{3}\left\vert 1\right\rangle _{A}\right) .
\end{equation}%
At this point, Alice has the atom $1$ and the cavity, while atoms $2$ and $3$
are sent to Bob. The state of the whole system composed by the two-level
atoms and the cavity mode field is
\begin{eqnarray}
\left\vert \psi \right\rangle _{total} &=&\frac{1}{2}\left[ \left\vert \Psi
^{+}\right\rangle _{1A}\left( C_{0}\left\vert g\right\rangle _{3}\left\vert
e\right\rangle _{2}+C_{1}\left\vert e\right\rangle _{3}\left\vert
g\right\rangle _{2}\right) \right.  \notag \\
&+&\left. \left\vert \Psi ^{-}\right\rangle _{1A}\left( C_{0}\left\vert
g\right\rangle _{3}\left\vert e\right\rangle _{2}-C_{1}\left\vert
e\right\rangle _{3}\left\vert g\right\rangle _{2}\right) \right.  \notag \\
&+&\left. \left\vert \Phi ^{+}\right\rangle _{1A}\left( C_{0}\left\vert
e\right\rangle _{3}\left\vert e\right\rangle _{2}-C_{1}\left\vert
g\right\rangle _{3}\left\vert g\right\rangle _{2}\right) \right.  \notag \\
&+&\left. \left\vert \Phi ^{-}\right\rangle _{1A}\left( C_{0}\left\vert
e\right\rangle _{3}\left\vert e\right\rangle _{2}+C_{1}\left\vert
g\right\rangle _{3}\left\vert g\right\rangle _{2}\right) \right] ,  \label{3}
\end{eqnarray}%
where, for convenience, we have defined the Bell states $\left\vert \Psi
^{\pm }\right\rangle _{1A}$ and $\left\vert \Phi ^{\pm }\right\rangle _{1A}$
as
\begin{equation}
\left\vert \Psi ^{\pm }\right\rangle _{1A}=\frac{1}{\sqrt{2}}\left(
-i\left\vert g\right\rangle _{1}\left\vert 1\right\rangle _{A}\pm \left\vert
e\right\rangle _{1}\left\vert 0\right\rangle _{A}\right) \text{,}
\end{equation}%
\begin{equation}
\left\vert \Phi ^{\pm }\right\rangle _{1A}=\frac{1}{\sqrt{2}}\left(
\left\vert g\right\rangle _{1}\left\vert 0\right\rangle _{A}\pm i\left\vert
e\right\rangle _{1}\left\vert 1\right\rangle _{A}\right) \text{.}
\end{equation}

As in the OP, the teleportation is completed after Alice measuring on
particle $1$ and cavity $A$ and sending her result to Bob, whom will know
which unitary operation to accomplish on its particles in order to recover
the entangled state that Alice wanted to teleport. Note, however, that
different from the OP, when comparing the teleported state resulting from
Eq. (\ref{3}) with the state to be teleported, Eq. (\ref{in}), we see that
partner $1$ was replaced by particle $3$. This explains the
\textquotedblleft partial teleportation\textquotedblright\ term used. The
experimental setup is shown, step by step, in Fig. \ref{scheme}. Next, we
show how Alice must proceed to perform the Bell state measurements.
\begin{figure}[t]
\includegraphics[width=4.5cm]{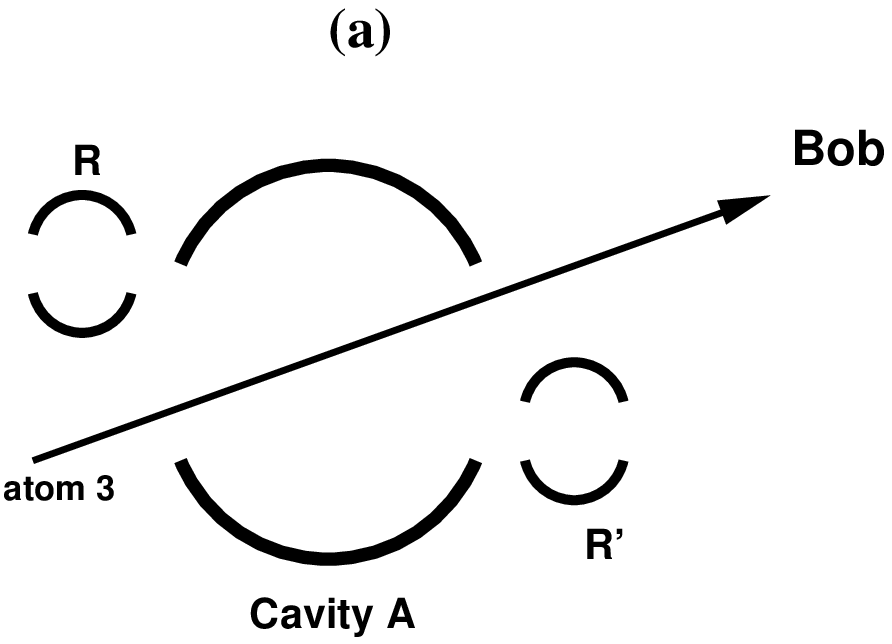}\newline
\vspace{0.5cm} \includegraphics[width=6cm]{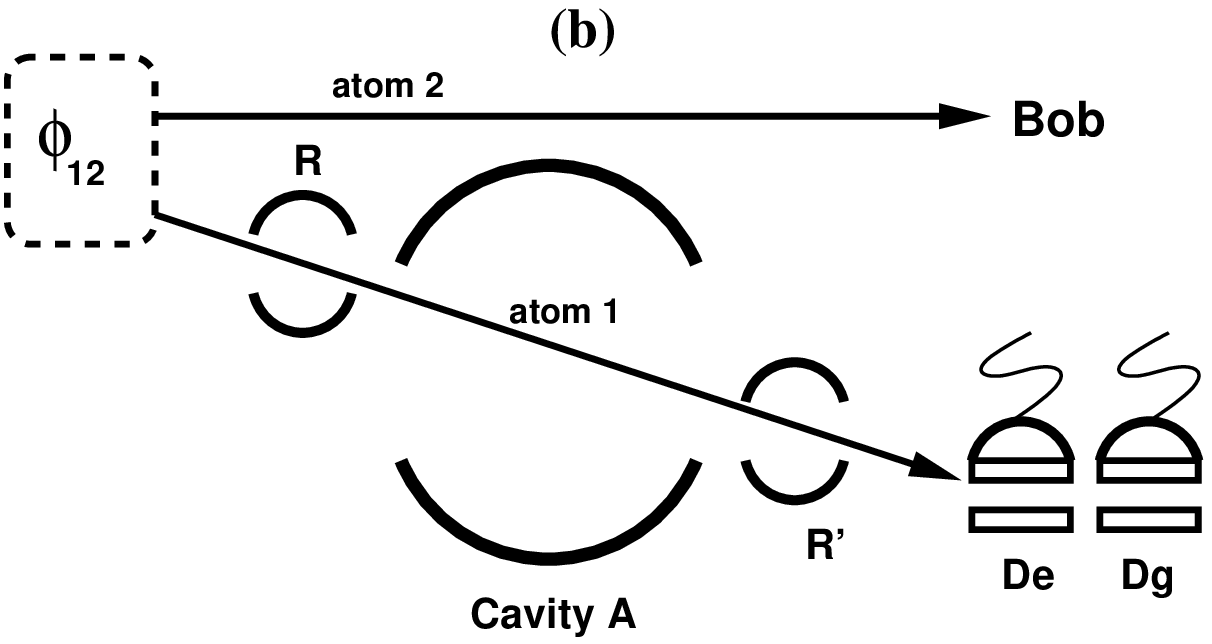}\newline
\vspace{0.5cm} \includegraphics[width=5cm]{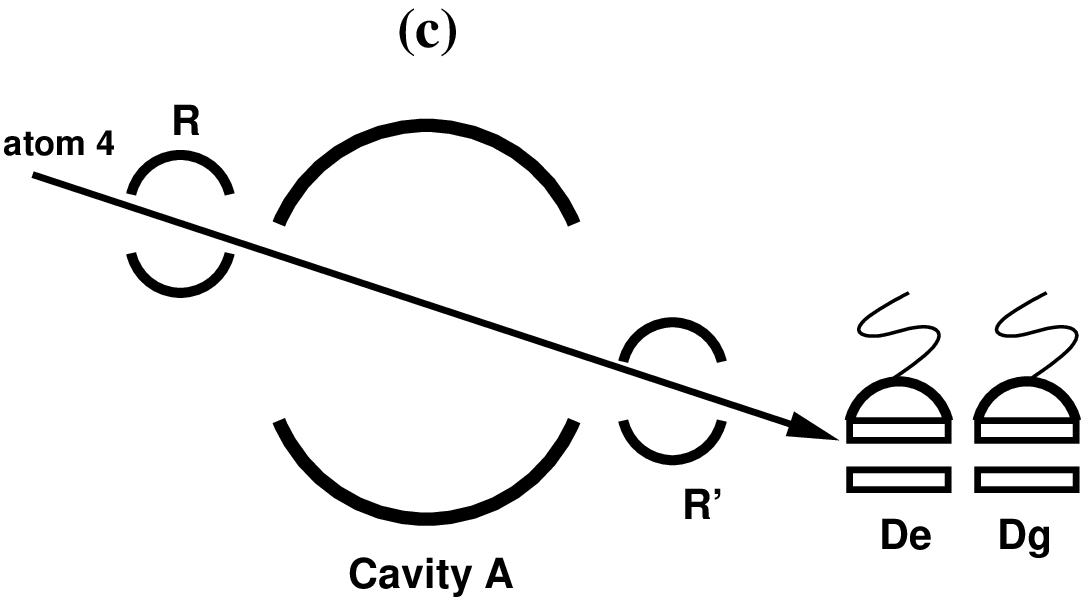}
\caption{Scheme for accomplishment of partial teleportation in cavity QED.
Three figures summarize the scheme: a) in a first step, the atom $3$
interacts resonantly with the cavity mode field $A$ and it is sent to Bob;
b) in a second step, the atom $2$ is sent rightly to Bob, while atom $1$
interacts 1) with a Ramsey zone $R$, 2) off-resonantly with the cavity mode $%
A$, and 3) with a Ramsey zone $R^{\prime }$, being detected in $D_{e}$
(excited) or $D_{g}$ (ground); c) in a third step, an auxiliary atom 4
interacts 1) with a Ramsey zone $R$ without suffering rotation, 2)
resonantly with the cavity field mode $A$, and 3) with Ramsey zone $%
R^{\prime },$ being detected in $D_{e}$ or $D_{g}$.}
\end{figure}

\subsection{Bell State Measurements}

First, atom $1$ crosses a Ramsey zone $R$, adjusted to produce the following
evolutions
\begin{equation}
\left\vert e\right\rangle \rightarrow \frac{1}{\sqrt{2}}\left( \left\vert
g\right\rangle +\left\vert e\right\rangle \right)   \label{ev1}
\end{equation}%
and
\begin{equation}
\left\vert g\right\rangle \rightarrow \frac{1}{\sqrt{2}}\left( \left\vert
g\right\rangle -\left\vert e\right\rangle \right) .  \label{ev2}
\end{equation}%
Then, atom $1$ crosses the cavity interacting off-resonantly with mode $A$,
with the interaction time adjusted to $\chi t=\pi $ (with $\chi =\lambda
^{2}/\delta $), resulting in the evolutions $\left\vert g\right\rangle
_{1}\left\vert 0\right\rangle _{A}\rightarrow \left\vert g\right\rangle
_{1}\left\vert 0\right\rangle _{A},$ $\left\vert g\right\rangle
_{1}\left\vert 1\right\rangle _{A}\rightarrow \left\vert g\right\rangle
_{1}\left\vert 1\right\rangle _{A},$ $\left\vert e\right\rangle
_{1}\left\vert 0\right\rangle _{A}\rightarrow \left\vert e\right\rangle
_{1}\left\vert 0\right\rangle _{A},$ and $\left\vert e\right\rangle
_{1}\left\vert 1\right\rangle _{A}\rightarrow -\left\vert e\right\rangle
_{1}\left\vert 1\right\rangle _{A}$. Next, atom $1$ crosses another Ramsey
zone $R^{\prime }$ adjusted like the Ramsey zone $R$ (see Eqs. (\ref{ev1}-%
\ref{ev2})). As a consequence, the states of the Bell basis evolve as
\begin{equation}
\left\vert \Psi ^{\pm }\right\rangle _{1A}\rightarrow \frac{1}{\sqrt{2}}%
\left\vert g\right\rangle _{1}\left( -i\left\vert 1\right\rangle _{A}\pm
\left\vert 0\right\rangle _{A}\right) ,
\end{equation}%
\begin{equation}
\left\vert \Phi ^{\pm }\right\rangle _{1A}\rightarrow \frac{1}{\sqrt{2}}%
\left\vert e\right\rangle _{1}\left( \left\vert 0\right\rangle _{A}\pm
i\left\vert 1\right\rangle _{A}\right) .
\end{equation}%
Fig. 1b shows the passage of the atoms 1 and 2 in the schematic setup. By
selective atomic state detection on atom $1$ it is possible to know if the
joint state is $\left\vert \Psi \right\rangle _{1A}$ or $\left\vert \Phi
\right\rangle _{1A}.$ Next, we have to discern the phases $(\pm )$ of the
Bell states. With the Ramsey zone $R$ turned off, another two-level atom
(atom $4$) in the ground state $\left\vert g\right\rangle _{4}$ is sent
through the cavity to interact resonantly with mode $A$ (see Fig. 1c) as
indicated by Eq. (\ref{hon}), with the interaction time $t=\pi /2\lambda $,
thus resulting in the following evolutions: $|g\rangle _{4}|0\rangle
_{A}\rightarrow |g\rangle _{4}|0\rangle _{A}$ and $|g\rangle _{4}|1\rangle
_{A}\rightarrow -i|e\rangle _{4}|0\rangle _{A}$. Next, the atom $4$ crosses
the Ramsey zone $R^{\prime }$ (according to Eqs. (\ref{ev1}-\ref{ev2})) such
that the Bell-states are written as
\begin{equation}
\left\vert \Psi ^{\pm }\right\rangle _{1A}\left\vert g\right\rangle
_{4}\rightarrow \left\{
\begin{array}{c}
\left\vert g\right\rangle _{1}\left\vert e\right\rangle _{4}\left\vert
0\right\rangle _{A}\text{ \ \ if (}+\text{)} \\
\left\vert g\right\rangle _{1}\left\vert g\right\rangle _{4}\left\vert
0\right\rangle _{A}\text{ if (}-\text{)}%
\end{array}%
\right. ,
\end{equation}%
\begin{equation}
\left\vert \Phi ^{\pm }\right\rangle _{1A}\left\vert g\right\rangle
_{4}\rightarrow \left\{
\begin{array}{c}
\left\vert e\right\rangle _{1}\left\vert g\right\rangle _{4}\left\vert
0\right\rangle _{A}\text{ \ \ if (}+\text{)} \\
\left\vert e\right\rangle _{1}\left\vert e\right\rangle _{4}\left\vert
0\right\rangle _{A}\text{ \ \ if (}-\text{)}%
\end{array}%
\right. \text{.}
\end{equation}%
Thus, by measuring atom $4$ we will be able to distinguish between the phase
$\left( -\right) $ or $\left( +\right) $. The perfect discrimination between
the four states composing the Bell base can be accomplished by Alice through
the detection of the atoms $1$ and $4$, separately. After that, Alice sends
a sign to Bob, whom accomplishes an appropriate rotation in the states of
the atoms $2$ and $3$ to complete the partial teleportation with $100\%$ of
fidelity and success probability, in the ideal case. The unitary operations
required by Bob are summarized in Table $1$.
\begin{table}[t]
\begin{tabular}[b]{||c||c||c||}
\hline\hline
BSM & $|\psi \rangle _{32}$ & Unitary operation \\ \hline\hline
$\left\vert \Psi ^{+}\right\rangle _{1A}$ & $C_{0}|g\rangle _{3}|e\rangle
_{2}+C_{1}|e\rangle _{3}|g\rangle _{2}$ & $\mathbb{I}_{3}\otimes \mathbb{I}%
_{2}$ \\ \hline
$\left\vert \Psi ^{-}\right\rangle _{1A}$ & $C_{0}|g\rangle _{3}|e\rangle
_{2}-C_{1}|e\rangle _{3}|g\rangle _{2}$ & $\sigma _{3z}\otimes \mathbb{I}%
_{2} $ \\ \hline
$\left\vert \Phi ^{+}\right\rangle _{1A}$ & $C_{0}|e\rangle _{3}|e\rangle
_{2}-C_{1}|g\rangle _{3}|g\rangle _{2}$ & $\sigma _{3y}\otimes \mathbb{I}%
_{2} $ \\ \hline
$\left\vert \Phi ^{-}\right\rangle _{1A}$ & $C_{0}|e\rangle _{3}|e\rangle
_{2}+C_{1}|g\rangle _{3}|g\rangle _{2}$ & $\sigma _{3x}\otimes \mathbb{I}%
_{2} $ \\ \hline\hline
\end{tabular}%
\caption{Results of the teleportation scheme. BSM denotes the resulting
measurement on atom $1$ and field mode $A$. Unitary operation denotes the
required operation by Bob after Alice communicating her results. The $%
\protect\sigma _{3j}$ is the Pauli operator $\protect\sigma _{j}$ acting on
atom $3$.}
\end{table}

\section{Decay of the free atomic excited state}

\subsection{Phenomenological operator approach (POA)}

Here we observe that the coupling of the atomic states to a surrounding
environment $\mathcal{E}$ can be described by the relations \cite{poa}
\begin{equation}
|g\rangle |\mathcal{E}\rangle \overset{U_{t}}{\longrightarrow }|g\rangle
\hat{\mathcal{T}}_{0}|\mathcal{E}\rangle ,  \label{1d}
\end{equation}%
\begin{equation}
|e\rangle |\mathcal{E}\rangle \overset{U_{t}}{\longrightarrow }|e\rangle
\hat{\mathcal{T}}_{e}^{\dagger }|\mathcal{E}\rangle +|g\rangle \hat{\mathcal{%
T}}_{g}^{\dagger }|\mathcal{E}\rangle \text{,}  \label{1e}
\end{equation}%
where $|\mathcal{E}\rangle $ denotes the initial state of the environment,
the operators $\hat{\mathcal{T}}$, acting on this state, account for the
atom-environment coupling, and $U_{t}$ denotes an unitary operation mixing
the atom to its environment. We will assume the environment $|\mathcal{E}%
\rangle $ in the vacuum state, which is an excellent approximation for
high-Q cavities in the microwave domain \cite{Brune96}. Accordingly, we
assume that $\hat{\mathcal{T}}_{0}^{\dagger }=\mathbf{1}$, $\hat{\mathcal{T}}%
_{e}^{\dagger }=\mathbf{f}(t)=e^{-\kappa t}\mathbf{1}$, $\hat{\mathcal{T}}%
_{g}^{\dagger }=\sum_{j}\mathbf{g}_{j}(t)\hat{b}_{j}^{\dagger }$, with $%
\sum_{j}|\mathbf{g}_{j}(t)|^{2}=1-e^{-2\kappa t}$, $\kappa $ denoting the
spontaneous decay rate of the atomic excited state, $\mathbf{1}$\ is the
identity operator, $b_{j}^{\dagger }$ is the creation operator, having a
corresponding annihilation operator $b_{j}$, of the $j$th oscillator mode of
the environment, and $t$ is the time elapsing after the atom suffering a
given excitation. With these assumptions, it is straightforward to verify
that the superposition $\left( \left\vert g\right\rangle +\left\vert
e\right\rangle \right) /\sqrt{2}$ leads to the reduced density operator
\begin{eqnarray}
\rho &=&\frac{1}{2}\left\{ \exp (-2\kappa t)|e\rangle \langle e|+\left[
2-\exp (-2\kappa t)\right] |g\rangle \langle g|\right.  \notag \\
&+&\left. \exp (-\kappa t)\left( |e\rangle \langle g|+|g\rangle \langle
e|\right) \right\} \text{.}  \label{1c}
\end{eqnarray}%
Note that the evolution (\ref{1d}) and (\ref{1e}) are consistent with the
well-known result that an unstable atomic state decays exponentially. In
this case, the phenomenological-operator evolution leads to the same atomic
density operator as the one we obtain using an \textit{ab-initio} master
equation approach. Moreover, due to recent advances in high-Q cavities \cite%
{harochenature}, we will neglect the damping time of the mode $A$.

\subsection{Decay of the teleported state}

To estimate the losses, we assume the whole state starting to decay after
the preparation of the quantum channel. In the first step, the
phenomenological operators used to introduce damping effects, Eqs. (\ref{1d}%
) and (\ref{1e}), are applied to the whole system until the time $t$. Then,
for each excitement suffered by the atoms during the teleportation process,
a new phenomenological operator is included, which modifies the decay
probability of the atomic states, and as a consequence, the fidelity of the
whole teleportation process. Summarizing the applications of the
phenomenological operators from the beginning, \textit{i.e.,} since the
preparation of the quantum channel until the end of our teleportation
protocol, which occurs at the instant that the fourth atom is detected, we
have to apply them soon after \textit{i)} the atom $1$ crossing the first
Ramsey zone ($t_{1}$); \textit{ii)} the atom $1$ crossing the second Ramsey
zone ($t_{2}$); \textit{iii)} the atom $4$ interacting resonantly with the
mode field cavity ($t_{3}$); \textit{iv)} the atom $4$ crossing the Ramsey
zone ($t_{4}$). After the inclusion of the decay via POA, the state of the
whole system by the time the teleportation is concluded becomes a mixture,
being represented by a reduced operator density when the reservoir is traced
out. In our estimative, we take the case of the teleported state in Bob
hands when Alice measures the Bell state $\left\vert \Psi ^{+}\right\rangle
_{1A}$ ($\left\vert g\right\rangle _{1}\left\vert e\right\rangle _{4}$). The
corresponding fidelity is shown in Fig. \ref{fid}. Note that at the time the
teleportation is completed ( $t_{4}$) the fidelity rounds $1$, indicating
that we can safely neglect losses occurring during the teleportation
process. In fact, taking $t_{1}=2\mu s$, as reported in \cite{RaimondRMP01},
we will have $t_{2}\simeq 5\times 10^{-4}s+2\mu s+t_{1}$, which is the
necessary time for the atom $1$ to interact dispersively with the cavity
field and to cross the Ramsey zone, $t_{3}=10^{-4}s+t_{2}$, which is the
necessary time for the atom-field resonant interaction, and $t_{4}=2\mu
s+t_{3}\simeq 6,06\times 10^{-4}s$, which is much shorter than the atomic
decay $\kappa ^{-1}\cong 10^{-2}s$, being the fidelity at this time $0.99$
as can be seen from Fig. \ref{fid}. However, as the time goes on, the decay
becomes faster and the fidelity is reduced to $2/3$ at the instant $%
t_{f}=5,78\times 10^{-3}s$. Therefore, the effective time during which the
teleported state is at our disposal for further operations is $t_{f}-t_{4}=$
$5,17\times 10^{-3}s$.
\begin{figure}[t]
\includegraphics{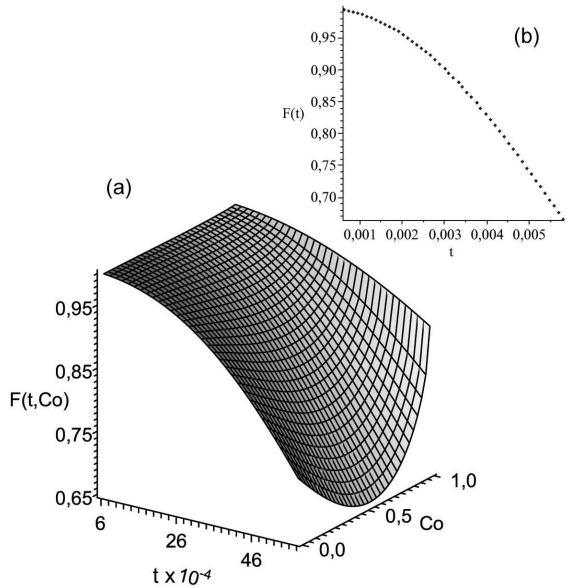}
\caption{Decay effects of the teleported state. In (a)The behavior of the
fidelity and its dependence with both the life-time of the atomic state and
the value of the coefficients of the state to be teleported. (b)The behavior
of the fidelity for the fixed values of the coefficients $C_{0}=C_{1}=1/%
\protect\sqrt{2}$.}
\end{figure}
.

\section{Fluctuation effect in the interaction time}

In this section we show that the fluctuation effect in the interaction time
due to the uncertainty in the atomic speed is not relevant as compared to
the decay effects of the teleported state presented in Section above. First,
we consider the impossibility of sharply fixing the atom-field interaction
time. The method adopted here is the same presented in Ref. \cite%
{MessinaEPJD02}. We introduce the probability density $f_{j}(t_{j};%
\widetilde{t_{j}})$, where $t$ is the true duration of the interaction
between the atom $j$ ($j=1,2$) and the cavity. We assume that $f_{j}(t_{j};%
\widetilde{t_{j}})$ is a Gaussian distribution centered around $\widetilde{%
t_{j}}$, e.g.,%
\begin{equation}
f_{j}(t_{j};\widetilde{t_{j}})=\frac{1}{\Delta _{j}\sqrt{2\pi }}\exp \left( -%
\frac{\left( t_{j}-\widetilde{t_{j}}\right) ^{2}}{2\Delta _{j}^{2}}\right) ,
\end{equation}%
where $\Delta _{j}=x\widetilde{t_{j}}$ and $x$ is a parameter related to the
uncertainty in the atomic velocity (around $0.5\%$ according to recent
experiments \cite{RauschenbeutelPRL99}), and therefore in the requested
interaction times $\widetilde{t_{j}}$. Thus, the density operator of the
system including the fluctuation effect is written as
\begin{equation}
\rho =\int \int \int_{-\infty }^{\infty }\left[ \prod%
\limits_{j=1}^{N}f_{j}(t_{j};\widetilde{t_{j}})dt_{j}\right] \left\vert \psi
\right\rangle _{total}\left\langle \psi \right\vert .
\end{equation}%
Here, for simplicity, we consider $N=1,2,3$ to describe the three
interactions between atoms $1$ and $4$ and the cavity. Following the steps
in Ref. \cite{MessinaEPJD02}, we obtain the fidelity given by
\begin{widetext}
\begin{eqnarray}
F&=&N^{2}\left[ 1/2\,{{C_{0}}}^{4} \left( {{\rm e}^{3/2\,{x}^{2}{\pi
}^{2}}}+{{\rm e} ^{1/2\,{x}^{2}{\pi }^{2}}}+2\,{{\rm e}^{{x}^{2}{\pi
}^{2}}} \right) { {\rm e}^{-3/2\,{x}^{2}{\pi }^{2}}}+ \left(
1-{{C_{0}}}^{2} \right)
 \left( 2-2\,{{C_{0}}}^{2} \right) \right. \nonumber \\
&&- \left. 2\,{{C_{0}}}^{2} \left( -{ {\rm e}^{1/2\,{x}^{2}{\pi
}^{2}}}-1+{{C_{0}}}^{2}{{\rm e}^{1/2\,{x}^{ 2}{\pi
}^{2}}}+{{C_{0}}}^{2} \right) {{\rm e}^{-3/4\,{x}^{2}{\pi }^{2 }}}
\right],
\end{eqnarray}
with
\begin{eqnarray}
N&=&\left( \sqrt {  2\,{{C_{0}}}^{2}{{\rm e}^{-1/2\,{x}^{2}{\pi
}^{2}}}+3- 2\,{{C_{0}}}^{2}-{{\rm e}^{-1/2\,{x}^{2}{\pi }^{2}}} }
\right) ^{-1}.
\end{eqnarray}
\end{widetext}The plot of the fidelity is shown in Fig. \ref{fvel}. Note
that the fidelity does not suffer a significant modification when
considering up to $3\%$ of uncertainty in the interaction time.
\begin{figure}[t]
\includegraphics[{width=7cm}]{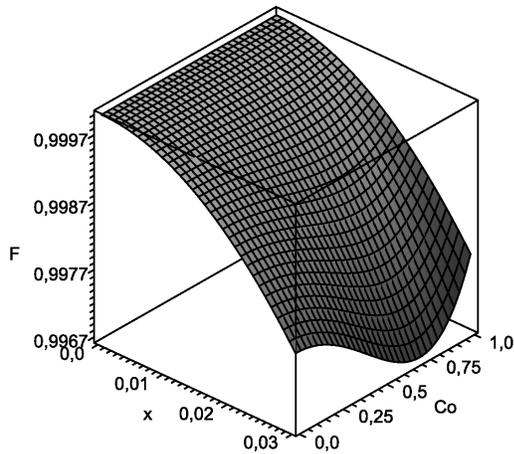}
\caption{Fidelity of the teleported state considering the fluctuation
effects in the atom-field interaction time. $C_{0}$ is the coefficient to be
teleported and $x$ is a parameter of uncertainty in the interaction time due
to the uncertainty in atomic velocities.}
\end{figure}

\section{Comments and Conclusion}

\qquad Since the teleportation protocol by Bennett \textit{et al}. \cite%
{Bennett93}, several other proposals have appeared, modifying slightly or
substantially the original protocol. In this paper we have explored a kind
of teleportation named by \textit{partial teleportation} \cite{LeePRA00}. In
our scheme, Alice has an atomic state to be teleported, given by an
entanglement of particles $1$ and $2$. Besides, Alice shares with Bob a
nonlocal channel composed by the joint state of a particle $A$ (represented
by a single mode of a high Q cavity) and a particle $3$ (represented by an
atomic state). After Alice performing a Bell measurement on the states of
particles $A$ and $1$, and informing Bob her result, the following
interesting result emerges, after the usual rotation by Bob: particle $3$
takes exactly the role of particle $1$ in the entanglement addressed to
Alice, but in Bob location. As the entanglement between the particles $1$
and $2$ is broken and a new entanglement between the particles $3$ and $2$
is created in a different place, this characterizes a partial teleportation.
Note that different from Ref. \cite{LuPLA00}, in our scheme the
teleportation occurs in only one particle of the entangled pair. To estimate
losses occurring during and after the teleportation process, we have used
the phenomenological operator approach (POA), as introduced in Ref. \cite%
{poa}. The fluctuation effect in the atom-field interaction time due to the
uncertainty in atomic velocities was also considered, showing a small
variation in the fidelity. Taking experimental parameters from recent
experiments in QED-cavity \cite{harochenature}, we believe that this scheme
can experimentally be accomplished using nowadays technology.

\subsection*{Acknowledgments}

We thanks CNPq and CAPES, Brazilian Agency, and PROPE/UCG - Universidade Cat%
\'{o}lica de Goi\'{a}s, for supporting this work.

\end{document}